\documentclass[journal]{IEEEtran}
\IEEEoverridecommandlockouts
\usepackage{url}
\usepackage{cite}
\usepackage{enumitem}
\usepackage{listings}
\usepackage[dvipsnames]{xcolor}
\usepackage{booktabs}
\usepackage{minted}
\usepackage{amsmath,amssymb,amsfonts}
\usepackage{algorithmic}
\usepackage[justification=centering]{caption}
\usepackage[table]{xcolor}
\usepackage{graphicx}
\usepackage{multirow}
\usepackage{textcomp}
\usepackage{booktabs}
\usepackage{multirow}
\usepackage{makecell}
\usepackage{graphicx} % for \resizebox
\usepackage{bbm}
\usepackage{threeparttable}
\usepackage{xcolor}

\def\BibTeX{{\rm B\kern-.05em{\sc i\kern-.025em b}\kern-.08em
    T\kern-.1667em\lower.7ex\hbox{E}\kern-.125emX}}
\usepackage{authblk}
    
\begin{document}
    
\title{SENSE: State-aware Emotion Navigation Storytelling Engine%
\thanks{\textcopyright~2026 IEEE. Personal use of this material is permitted. Permission from IEEE must be obtained for all other uses, in any current or future media, including reprinting/republishing this material for advertising or promotional purposes, creating new collective works, for resale or redistribution to servers or lists, or reuse of any copyrighted component of this work in other works. This work has been published in IEEE Transactions on Games. DOI: 10.1109/TG.2026.3733998}}

\author[1,*]{Yi Xia}
\author[1]{Pablo Carrasco Velo}
\author[1]{Mudit Paliwal}
\author[2]{Ibrahim Khan}
\author[1]{Yifan Geng}
\author[3]{Mustafa Can Gursesli}
\author[3]{Juho Hamari}
\author[2]{Ruck Thawonmas}

\affil[1]{Graduate School of Information Science and Engineering, Ritsumeikan University, Ibaraki, Japan\\
\texttt{yi.xia@ice.ci.ritsumei.ac.jp}, \texttt{\{gr0797xi, gr0798hh, gr0797rx\}@ed.ritsumei.ac.jp}}
\affil[2]{College of Information Science and Engineering, Ritsumeikan University, Ibaraki, Japan\\
\texttt{khan@fc.ritsumei.ac.jp}, \texttt{ruck@is.ritsumei.ac.jp}}
\affil[3]{Faculty of Information Technology and Communication Sciences, Tampere University, Tampere, Finland\\
\texttt{\{can.gursesli, juho.hamari\}@tuni.fi}}
\affil[*]{Corresponding author: \texttt{yi.xia@ice.ci.ritsumei.ac.jp}}

\IEEEaftertitletext{\vspace{-2em}}
\maketitle
\begin{abstract}
This paper presents SENSE, a state-aware framework for generating playable branching visual novels with multi-track emotional navigation. Integrating a state-based narrative architecture called MIND, a structure analyzer, and a path-aware context management module, SENSE produces narratives that are both structurally coherent and emotionally rich. From minimal high-level inputs, it generates multiple intersecting routes while preserving character consistency and narrative causality. Evaluations using LLM judges, affective metrics, and visual assessments indicate SENSE outperforms baselines in narrative diversity and robust asset integration, while preliminary human trials show directional improvements in emotional fidelity alongside comparable enjoyment.
\end{abstract}


\section{Introduction}

Visual novels (VNs) provide deeply interactive and immersive experiences by integrating narrative text, visual presentation, and branching choices. In such narrative-driven game design, emotional arcs are considered the logical core guiding plot progression and fostering player resonance~\cite{reagan2016emotional, wen2025all}. Although hierarchical ``Outline-to-Detail" (O2D) paradigms have streamlined scriptwriting efficiency, large language models (LLMs) struggle with complex networked topologies~\cite{10.1145/3746027.3755541}. The core challenge lies in the systematic graph-level failures LLMs encounter during long-term navigation, including missing passages, trap-like cycles, and causal asymmetries. Following that, a recent study by Puchalski et al., \cite{sym18010113} attempted to rectify these defects through post-hoc, name-based repair loops. However, such approaches primarily focus on the physical path coherence and do not address the guidance of emotional intent. Without emotional orchestration, even a physically intact path is prone to emotional drift~\cite{wen2025all}, causing the player’s experience to deviate from the intended emotional trajectory.  

In order to address these challenges, we introduce the State-aware Emotion Navigation Storytelling Engine (SENSE), a framework for scalable, emotionally salient, and visually aligned VN generation. SENSE decouples creative narrative logic from deterministic code, translating LLM-generated scripts into executable Ren’Py code while isolating path-dependent histories to prevent cross-branch interference. Its Memory and Intelligent Navigation Director (MIND) incorporates the Structural Thematic and Adaptive Router (STAR) to maintain thematic coherence while allowing branching divergence. State-based narrative mediation integrates historical path discrepancies into coherent psychological narratives, transforming abrupt emotional shifts into logical, emotionally resonant progressions. This state-aware, topology-validated architecture is designed to support structurally sound branching while mitigating emotional drift across routes.

The main contributions of this study are as follows:
\begin{itemize}
    \item \textbf{SENSE: a state-aware emotion navigation storytelling engine.} 
    We propose SENSE, an end-to-end engine that generates playable VNs from minimal inputs by explicitly modeling narrative state, branching history, player choices, and emotional trajectories.

    \item \textbf{Emotion-guided branching synthesis via MIND.} 
    The MIND state-based architecture integrates macro-level emotional planning with path-aware script generation, guiding emotional progression while supporting causal consistency across branching narratives. 

    \item \textbf{Deterministic topology validation and context-aware script synthesis.} SENSE supports structurally sound branching narratives through a mechanism called Structure Analyzer, which corrects node errors, cycles, and orphans. It further generates route-specific scripts via topology-driven, path-aware synthesis, improving context consistency across shared nodes.

    \item \textbf{Executable integration with context-aware assets.} 
    We compile validated scripts and visuals into playable VNs, enhancing semantic fidelity and interactive coherence.
\end{itemize}

\section{Related Work}
\subsection{Emotional Arcs and Narrative Structure and Generation}
Stories often follow predictable emotional trajectories, a concept rooted in early narrative theory and validated by sentiment analysis. Reagan et al.~\cite{reagan2016emotional} showed that these arcs cluster into six fundamental shapes recurring across cultures and genres. Maintaining emotional trajectories is critical in interactive media, where player choices introduce variability. Frameworks such as PACE~\cite{hernandez2015keeping} use automated planning to keep players on target emotional curves, while systems like the Virtual Storyteller~\cite{theune2003virtual} generate story plots influenced by autonomous character actions, enabling the narrative affect to align with user preferences. While preliminary work, such as Wen et al.~\cite{wen2025all}, applies basic emotional patterns to simplified branching games, integrating emotional trajectories into LLM-based generation of narrative-driven VNs with long texts and intersecting branches remains underexplored.

Hierarchical narrative generation, separating high-level plot planning from surface text realization, is a well-established paradigm in computational storytelling. Early symbolic systems, such as TAIL-SPIN~\cite{meehan1977tale}, demonstrated the feasibility of top-down narrative construction. Modern neural approaches decompose generation into event-to-event and event-to-sentence levels~\cite{martin2018event}, while Fan et al.~\cite{fan2018hierarchical} generate narrative sketches and expand them into full stories, improving coherence. Recent methods incorporate dynamic outlining to adjust high-level plans based on intermediate content~\cite{wang2025generating}, addressing the rigidity of static hierarchical approaches. PlotMachines~\cite{rashkin2020plotmachines} treats story outlines as sets of guiding phrases for generation. Together, these approaches illustrate the evolution of hierarchical strategies in automated storytelling. These hierarchical approaches mirror human writing and help mitigate long-range dependency issues, providing conceptual support for our SENSE pipeline that separates narrative planning, state-aware script synthesis, and executable integration.

\subsection{LLM-Based Branching Narrative Generation}

LLMs have catalyzed new approaches to interactive narrative generation, producing branching stories via few-shot prompting and in-context learning. Recent systems such as ``what-if" demonstrate that LLMs can generate and visualize branching narratives, enabling authors to explore ``what-if" scenarios through mixed-initiative interfaces~\cite{huang2024if}. GENEVA~\cite{leandro2024geneva} generates complete branching narratives from high-level descriptions and renders them as interactive graphs that authors can refine. TaleFrame~\cite{wang2025taleframe} deploys locally-hosted LLMs fine-tuned on preference datasets to achieve fine-grained control over story generation. Despite these advances, LLM-based approaches still struggle with global consistency across branching paths~\cite{rameshkumar2025reasoning, sym18010113}, producing structural errors such as edge omission, branch pruning, and hallucinated connections, which motivates the need for topology-aware validation and state-based narrative management.

\subsection{LLM-driven Multi-modal Narrative Generation}

Recent studies show that LLMs can mimic convincing human social behaviors~\cite{park2023generative}, inspiring multi-agent systems like AutoGen~\cite{wu2024autogen}, CAMEL~\cite{li2023camel}, and MetaGPT~\cite{hong2024metagpt} that provide fundamental role-playing and coordination capabilities. Building on this, multi-agent debate protocols~\cite{du2024improving,liang2024encouraging} and collaborative narrative frameworks like IBSEN~\cite{wang2024ibsen}, Agents’ Room~\cite{micaloni2024agents}, and PatternTeller~\cite{delima2024patternteller} leverage structured interactions. Despite structural gains, current hierarchical frameworks, such as~\cite{10.1145/3746027.3755541}, still have room for improvement in text-image alignment. Inspired by multi-agent paradigms, we adopt a role-specialized, agent-based workflow, in which distinct generative and evaluative components are orchestrated through an iterative multi-modal selection and validation loop to improve visual–textual alignment.

\section{Methodology}
\subsection{System Overview}
\label{subsec:system_overview}

\begin{figure*}[tbp]
    \centering
    \includegraphics[width=\textwidth]{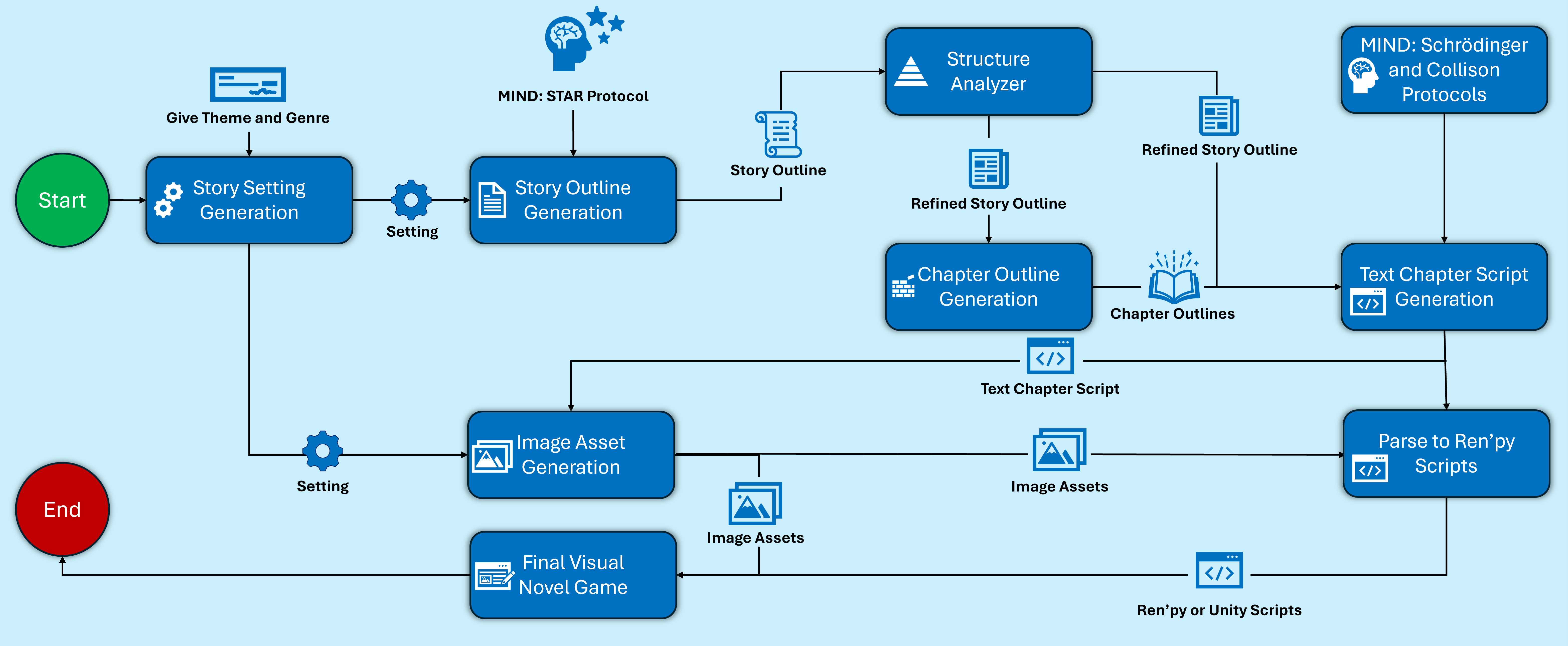}
    \caption{Overall architecture of the SENSE system.}
    \label{fig:sense}
\end{figure*}

To fulfill the dual requirements of scalable narrative expansion and emotional expressiveness, SENSE is designed as an integrated end-to-end framework driven by a structural emotional blueprint and hierarchical narrative-to-code decoupling. As illustrated in Figure \ref{fig:sense}, the system autonomously generates playable VNs from minimal high-level inputs (theme, genre, and emotional arc type). It achieves this by anchoring the LLM's creative synthesis to dynamic affective trajectories, while strictly separating the narrative prose from deterministic logical routing and code compilation. Audio generation is explicitly excluded from this study to limit modality scope, while retaining textual and visual asset generation.

The system follows seven decoupled phases, integrating MIND with deterministic structure analysis and path-aware context management:

\begin{enumerate}[label=\alph*)]
    \item The system expands the minimal inputs into a comprehensive world setting.
    \item MIND utilizes its internal STAR Protocol to generate a macroscopic story outline embedded with emotional trajectory tags.
    \item A deterministic safeguard mechanism called the Structure Analyzer processes this outline to detect and resolve logical deadlocks and output a valid routing topology. 
    \item The system derives granular chapter outlines from the story structure processed by the Structure Analyzer.
    \item MIND internalizes path-dependent histories through state-based narrative mediation to generate engine-agnostic, tag-annotated intermediate chapter scripts.
    \item The pipeline produces context-aware image assets derived from the world setting and chapter scripts.
    \item A deterministic parser integrates these visual assets with the tagged scripts to synthesize executable engine code, ensuring robust syntactic validity.
\end{enumerate}

\subsection{The MIND state-based Architecture}
\label{subsec:mind_architecture}

The MIND module serves as the inference-time orchestration layer and state-based narrative controller of SENSE, bridging the gap between minimal input and granular narrative execution. Rather than modifying the underlying LLM architecture, MIND orchestrates inference through structured prompt-based constraints and dynamic context injection. Following that,  to support emotional trajectory guidance and logical consistency, MIND operates through a dual-layered governance strategy: macroscopic planning via the internal STAR Protocol and microscopic synthesis through Path-Aware Narrative Synthesis.

\subsubsection{Macroscopic Governance: The STAR Protocol}
In the initial phase, MIND employs the STAR Protocol to establish a foundation for thematically coherent narrative expansion. This protocol is operationalized through prompt-based constraint injection, wherein the LLM is furnished with a strict logical framework embedded within its system instructions. The protocol implements a dual-governance strategy to regulate the global emotional trajectory. The \textit{Spine}, or primary narrative path, is governed by hard constraints: MIND retrospectively derives narrative logic to conform to the given emotional arc. In contrast, the \textit{Limbs}, or branching paths, are governed by autonomous reasoning, wherein emotional trajectories are dynamically generated through the Reaction instruction specified in the system prompt. This instruction orchestrates the causal interplay between internal character factors (e.g., psychological states and value systems) and external contextual factors (e.g., environmental pressures and resource scarcity). Based on these emotional trajectories, explicit emotional tags (i.e., Rise or Fall) are assigned to each narrative node, thereby enabling the STAR Protocol to construct a structured emotional blueprint that encodes predetermined outcome constraints. This mechanism supports global emotional variation across narrative branches, while mitigating the risk of model-induced affective monotony.

\subsubsection{Microscopic Logic Safety: The Schrödinger and Collision Protocols}\label{subsec: microscopic_safety}

To mitigate logical hallucinations during branching synthesis, MIND incorporates two specialized execution modes defined within the script generation prompts. The activation of these modes is strictly controlled by the system's underlying code, which dynamically constructs the user prompt based on narrative routing conditions.

When the code's routing logic detects structural ambiguity (specifically, when multiple diverging player choices route to the same subsequent chapter), the system conceptualizes this convergence as a \textit{narrative superposition}. To handle this state where the actual player choice remains unobserved, it dynamically injects ``POSSIBLE PLAYER CHOICES'' into the user prompt, triggering \textit{Mode A (The Schrödinger Rule)}. Named after the quantum mechanical thought experiment, this rule dictates that the generated narrative must remain valid across all potential player histories without collapsing into a logical contradiction. To maintain this consistency before writing the merged scene, this triggered rule enforces a protocol across three dimensions:
\begin{itemize}
    \item \textbf{Physical Check}: If physical states (e.g., item possession or health) are in conflict, the system must not describe the specific item or body part, focusing instead on the environment.
    \item \textbf{Relational Check}: If past interactions conflict, NPC reactions must be described as ``Unreadable,'' ``Complex,'' or ``Silent'' to ensure sentiment validity.
    \item \textbf{Informational Check}: If the player's knowledge state is ambiguous, the narrative focus shifts to the overarching mystery or goal rather than specific solutions.
\end{itemize}
If any of these dimensions remain in conflict, the prompt makes the LLM shift to an internal monologue about general consequences or introduces a sudden external event to drive the story forward without logical friction.

Conversely, \textit{Mode B (The Collision Rule)} is activated when the routing algorithm identifies a unique branch or ending. This acts as the narrative observation that collapses the superposition. The code dynamically constructs a user prompt that explicitly states ``The player DEFINITELY chose: [Action]'' or contains a ``CRITICAL INSTRUCTION''. This mode completely ignores ambiguity and enforces \textit{Specific Causality}. The system must explicitly reference the specific choice (e.g., describing blood on a weapon if the player chose to attack, or shame if they chose to run). By colliding the chapter outline with the specific choice's tone and outcome, MIND creates a unique divergence that is logically robust and responsive to player choices.

\subsection{ Structure Analyzer}
During the macroscopic narrative planning phase, specifically the outline generation, LLMs often struggle with strict graph-theoretic consistency. We find that even when adopting a baseline-inspired two-step generation pipeline (first drafting the raw narrative events in natural language and subsequently converting them into a structured JSON format), the LLM still frequently hallucinates node identifiers (IDs) or generates illogical structural loops. To address this limitation, SENSE introduces a deterministic safeguard mechanism called the Structure Analyzer, which consolidates the story graph before chapters and scripts synthesis.

This mechanism operates through three core deterministic processes:

\begin{itemize}
    \item \textbf{Heuristic Priority-Based Node Resolution:} LLMs frequently mismatch numerical chapter IDs during the JSON conversion step while accurately retaining semantic titles. To rectify broken transition links, the analyzer employs a multi-dimensional target resolution algorithm. Instead of relying solely on exact ID matching, it evaluates potential target nodes through a strict priority sequence: Full Title Match, Explicit ID Match, Cleaned Title Match (filtering out ID and emotion tags), and finally, a composite Branch and Keyword overlap. This heuristic approach supports robust edge reconstruction even when the LLM's numerical indexing fails.
    
    \item \textbf{Cycle Detection and Heuristic Breaking:} Non-linear narrative generation is highly susceptible to structural cycles (e.g., self-loops or complex backward loops) that can cause infinite rendering errors. The system utilizes a Depth-First Search (DFS) algorithm to detect all topological cycles. Rather than randomly severing connections, it breaks cycles intelligently by computing a \textit{Connection Quality Score} for each edge involved in a loop. This heuristic scoring system incorporates narrative domain knowledge, rewarding logical transitions, such as natural index progression or connections within the same branch designation, while heavily penalizing structural regressions. By evaluating these scores, the algorithm deterministically isolates and breaks the weakest link, thereby preserving the most logical narrative flow.
    
    \item \textbf{Generalized Orphan Resolution and Topological Preservation:} In the final stage, the algorithm systematically distinguishes between \textit{true orphans} (nodes completely lacking incoming routes) and \textit{falsely flagged orphans} (disconnected sub-graphs retaining internal parent-child links). To rescue the latter, the system executes a global reverse-validation of all narrative pointers. By evaluating implicit connections and employing type-agnostic identifier comparisons to mitigate data type hallucinations (e.g., integer-string confusion), the algorithm dynamically reconstructs missing edges and updates the traversal maps. Ultimately, only confirmed true orphans are pruned. The system subsequently utilizes this fully validated graph structure to autonomously rectify the original narrative outline, yielding a structurally validated foundation for subsequent script synthesis.
\end{itemize}

\subsection{Topology-Driven Context Injection and Script Synthesis}
\label{subsec:script_synthesis}
SENSE programmatically activates microscopic logic protocols by evaluating the structural properties of target nodes and dynamically constructing prompt payloads. This mechanism is designed to provide the generative model with specific contextual triggers based on the graph's topological identity.
\begin{itemize}
    \item \textbf{Schrödinger State Maintenance (Convergence Logic):} For chapters with a topological in-degree $\ge 2$ representing narrative convergence, the system aggregates all potential player choices, including their labels and underlying router logic, into the prompt payload under the flag \texttt{POSSIBLE PLAYER CHOICES (SCHRÖDINGER STATE)}. This serves as the deterministic trigger for \textit{Mode A (The Schrödinger Rule)}, established in Section \ref{subsec: microscopic_safety}.
    
    \item \textbf{Collision and Ending Divergence (Divergence Logic):} For terminal nodes identified in the global Identity Map, the system supports ending divergence by generating unique variant files (e.g., \texttt{\_End\_xx}). By declaring a definitive player action in the prompt, the system triggers Mode B (The Collision Rule). This collapses the narrative superposition, requiring the model to detail the specific consequences associated with that singular choice.
    
    \item \textbf{Path-Aware Context and Sequential Synthesis:} SENSE synthesizes content sequentially along individual routes, creating unique context signatures for shared chapters conditioned on their ancestor sequences. The system builds a Path-Aware Context by retrieving the full textual transcripts of the current path. This allows shared topological nodes to produce distinct script variants, isolating narrative causality and avoiding inter-route interference.
\end{itemize}

\subsection{VN Image Assets Generation}
We propose an automated, identity-preserving generation pipeline  (see Fig. S1 in Section S7 of the supplementary file in the project repository; see Section IV-F for access information)
 designed to synthesize emotionally aligned visual assets while mitigating character drift. In contrast to the baseline approach, where the LLM generates novel prompts without contextual awareness of prior iterations~\cite{10.1145/3746027.3755541}, which often exacerbates identity instability, our method relies on the deterministic extraction of structured visual attributes from the world-setting configuration to establish a strict Base Identity. For narrative backgrounds, transient visual cues parsed from chapter scripts are systematically merged with the parent location's metadata to maintain contextual grounding.

To generate dynamic character expressions while minimizing identity degradation, the pipeline employs an image-to-image approach. An LLM maps abstract narrative tags (e.g., \textit{[anger]}) to the closest match within a 34-emotion dictionary derived from the Ekman and GO frameworks~\cite{ekman1992argument, cowen2017self} to extract corresponding facial descriptors for prompt construction. Crucially, an IP-Adapter~\cite{ye2023ip-adapter} enforces strong semantic conditioning directly from the Base Identity. In parallel, an automated object detection algorithm (YOLOv8)~\cite{yolov8_ultralytics} isolates and inpaints solely the facial region. This architecture ensures that emotional variations are applied exclusively to facial features without altering the character's core structure or attire.

In order to regulate output quality and prevent character drift, the pipeline utilizes a vision-language model (VLM) within an iterative generative-discriminative loop. During each iteration, the generative model outputs a batch of candidate images ($N=3$). The VLM evaluates these candidates against predefined quality criteria, which specifically include verifying strict consistency with the Base Identity alongside general visual fidelity. If the best-scoring image fails to satisfy the criteria, the VLM generates textual correction feedback. This feedback is appended to the prompt for the subsequent iteration. The cycle repeats until the image passes validation or reaches the maximum iteration limit ($T_{max}=2$).

\subsection{Script Integration and Backend Generation}\label{sec: backend}

This module bridges the gap between the intermediate text-based VNs script and executable runtime artifacts by generating engine-specific outputs for different backends. It operates on a shared script representation and applies a unified script-level integration strategy, enabling the same narrative content to be deployed across multiple execution environments. The current implementation is demonstrated using a Ren'Py integration. The intermediate representation is reusable, and similar conversion logic can be applied when adapting the system to other platforms, such as Unity, with platform-specific integration stages.

\subsection{Game script conversion and asset integration}
This stage transforms the intermediate text-based VNs script into an executable Ren'Py script, binding all pre-generated visual assets to their corresponding narrative elements. For this stage, we developed a series of parsers utilizing regular expressions to ensure that the resulting script can be executed directly without manual intervention, achieving high accuracy in the text-to-game conversion process.

In this stage, two distinct parsers operate sequentially to achieve this conversion. The first parser integrates previously generated images into the chapter structure. To facilitate asset integration, it identifies labels in the script and assigns asset identifiers, creating the links required for further conversion into a game script. The second parser processes the script generated by the first parser into a playable Ren'Py game. 

This script begins by initializing all assets within the Ren'Py ``script" file, a necessary step enabling Ren'Py to display images based on asset names. Additionally, this script implements core game mechanics, including interactions and decision points where the narrative branches into different routes. It utilizes the background labels provided by the processed script to display the correct background image and matches character names from dialogues to their corresponding asset images for character display. To prevent character overlap, a queuing mechanism limits the display to a maximum of three characters at any given time, positioned on the left, right, and center of the screen. When a fourth character with dialogue is encountered, the script replaces the oldest character image on screen, implementing a queue-based replacement strategy.

\section{Experiments}
\subsection{Experimental Setup}

\subsubsection{Frameworks and Generation Setup}
We evaluate the proposed SENSE framework against two frameworks: (1) Baseline: the recent hierarchical VN generation framework by Zhang et al., which lacks explicit affective modeling~\cite{10.1145/3746027.3755541}; and (2) WithoutMIND: an ablated SENSE pipeline stripped of the MIND architecture. We selected ``Climate Change" as the central theme for all 54 generated VNs, spanning three distinct genres with 6 VNs per framework. These three genres were specifically selected because they are distantly related and heterogeneous, ensuring a rigorous test of the framework's generalization capability across diverse narrative spaces. To standardize evaluation across frameworks, we instructed the LLM during outline generation to target a maximum of 6 chapters per narrative route. Despite LLMs' inherent variance in strictly following numerical limits, this soft constraint effectively bounded the narrative scale, ensuring comparable lengths for fair cross-modal assessment. To verify SENSE's affective saliency, we assigned one of six distinct emotional trajectories to each of the 6 VNs within every genre for SENSE frameworks, whereas the baseline operated without explicit constraints. 

The choice of climate change as the main theme is strategically grounded in its dual nature as both a complex and multifaceted global issue requiring exceptional narrative complexity and a representative application for Serious VNs\cite{Gursesli16022026}. By demanding high logical coherence and precise emotional pacing to effectively engage players and foster social awareness, this theme provides a rigorous testbed for our MIND mechanism's ability to balance structural constraints with prose rendering. Furthermore, utilizing this theme ensures our results provide a robust foundation for future extensions into the narrative experience and efficacy of AI-generated stories in serious game contexts.

\subsubsection{Comparative Integrity and Route Extraction Protocol}\label{subsub:Comparative_Integrity}
To ensure a controlled comparison, all frameworks utilized GPT-5-mini as the underlying LLM for text and script generation. Aside from this model alignment, the baseline was executed according to its original default settings. During preliminary testing on the baseline, we observed instances of technical limitations, such as outline inaccuracies and syntax bottlenecks. Although its script validation mechanism is designed to perform error correction using Ren’Py lint feedback, the authors acknowledge it resolves approximately 76\% of syntax errors rather than eliminating them entirely~\cite{10.1145/3746027.3755541}. Following the original implementation, we limited this automated correction process to three iterations.

We developed a dedicated parsing script grounded in Ren'Py syntax to isolate storytelling quality from code execution failures. This script traverses the branching logic of all methods to reconstruct player-facing text into pure-text narrative routes. This protocol acts as a lenient extraction for the baseline by bypassing non-fatal execution errors to salvage structurally readable content. This approach evaluates the baseline at its peak narrative performance rather than penalizing its architectural limitations.

\subsection{Primary Evaluation: LLM-as-a-Judge Protocol}\label{subsec: judge_protocol}

Conventional metrics like BLEU and ROUGE rely primarily on surface-level lexical overlap. However, such metrics often fail to capture the structural and creative dimensions crucial for evaluating open-ended storytelling. Consequently, we adopt the LLM-as-a-Judge paradigm as a validated and scalable proxy for narrative assessment. Following the evaluative framework established by Chhun et al.~\cite{10.1162/tacl_a_00689} and subsequently adopted in~\cite{zhang2025apbased}, we assess the generated results across six orthogonal criteria: Relevance (RE), Coherence (CH), Empathy (EM), Surprise (SU), Engagement (EG), and Complexity (CX). To ensure evaluative consistency, we conduct three independent evaluation trials for each narrative route at a zero-temperature setting ($T=0$) using distinct random seeds. The resulting scores, measured on a 5-point Likert scale, are subsequently aggregated into an Overall (OV) quality metric to mitigate the impact of stochastic variability. This holistic methodology serves to quantify the comparative performance of the SENSE framework against the baselines across diverse narrative dimensions.

Given that each VNs case yields multiple narrative routes, we conduct route-level evaluations using two distinct sampling pools: \textit{AllMean}, which incorporates the complete distribution of successfully generated routes to reflect system-wide expected quality, and \textit{Top-3} (or all, if $N < 3$), which isolates the highest-scoring routes per case to assess upper-bound creative potential. Evaluating this Top-3 subset is statistically imperative for an equitable comparison. Baseline methods frequently produce malformed scripts, resulting in severely restricted route yields. Consequently, relying solely on the AllMean pool disproportionately penalizes high-yield generative frameworks. A baseline may achieve a deceptively high mean from isolated serendipitous successes, whereas a robust system's comprehensive distribution inherently includes average-quality routes that dilute the aggregate score. By adopting the Top-3 (``Best-of-$N$'') alignment, we explicitly control for these sample size discrepancies, mitigating statistical dilution and enabling a direct comparison of the frameworks' upper-bound creative capabilities.

\subsection{Quantitative Analysis of Affective Salience}\label{subsec:second_eval}

Complementing the primary evaluation of story-level narrative quality, we conduct a fine-grained affective analysis to investigate the framework's capacity to enhance emotional intensity and affective richness. This evaluation utilizes the DeBERTa-based model of Christ et al. (2024), which is specifically optimized for modeling continuous valence ($V$) and arousal ($A$) trajectories in narratives using a context-aware sliding window with a radius of four sentences to integrate surrounding emotional context~\cite{christ-etal-2024-modeling}. The model demonstrates strong empirical reliability, with reported Concordance Correlation Coefficients (CCC) of 0.8221 for valence and 0.7125 for arousal.

Following the conceptualization of affective salience in~\cite{revers2025dissociating}, which defines it as the valence distance from neutral in either direction, we adopt this conceptualization to quantify emotional depth through a two-stage analytical pipeline. For each sentence $i (i = 1, ..., n)$, we compute its affective deviation $|E_i - 0.5|$ and identify its salience status based on a predefined non-salient range of $[0.35, 0.65]$, where $E \in {V, A}$ denotes the affective dimensions of valence and arousal, respectively. These sentence-level values are subsequently aggregated into two route-level metrics:
\begin{itemize}
\item \textbf{Affective Salience (AS):} quantifies the average intensity of emotional expression by computing the mean absolute deviation of all sentences in a route from the neutral baseline of $0.5$, effectively measuring the narrative's overall emotional intensity:
\begin{equation}
AS = \frac{1}{n} \sum_{i=1}^{n} |E_i - 0.5| \label{eq:as}
\end{equation}
\item \textbf{High-Salience Proportion (HSP):} identifies affective richness by calculating the percentage of sentences whose valence or arousal values fall outside the non-salient range:
\begin{equation}
HSP = \frac{1}{n} \sum_{i=1}^{n} \mathbbm{1}(E_i > 0.65 \lor E_i < 0.35) \times 100\% \label{eq:hsp}
\end{equation}
where $\mathbbm{1}(\cdot)$ is the indicator function that outputs $1$ if the affective value $E_i$ (where $E \in \{V, A\}$) is salient and $0$ otherwise.
\end{itemize}

\subsection{Visual Evaluation: Cross-Modal Alignment and Character Identity Preservation}\label{subsec:visual_evaluation}

To evaluate the quality of the generated visual assets, we assess both cross-modal text-image alignment and character identity preservation. Following Lin et al.~\cite{10.1007/978-3-031-72673-6_20}, we adopt VQAScore via the CLIP-FlanT5-XXL model as one of our metrics to quantify prompt alignment, computed as the probability of answering ``Yes'' to ``Does this figure show {prompt}?''. To directly measure cross-expression character persistence, we adopt two image-to-image similarity metrics from the evaluation protocol introduced in~\cite{ruiz2023dreambooth}: DINO and CLIP-I. Specifically, we compute the cosine similarity between the embeddings of each generated expression variant and its corresponding base character reference image, using self-supervised ViT-S/16 DINO and CLIP visual encoders, respectively. Together with VQAScore, these metrics evaluate whether SENSE preserves robust character identity across emotional variations while faithfully capturing prompt semantics.

\subsection{Statistical Methodology and Significance Testing}\label{subsec:statistical_m}
To evaluate statistical significance across textual and visual dimensions, we implement an adaptive pipeline determined by the Shapiro-Wilk normality test. For overall evaluation across all experimental groups, we employ a one-way ANOVA or the Kruskal-Wallis H test as an omnibus measure to identify overall differences. To isolate the contribution of specific modules, we conduct pairwise comparisons using Welch’s $t$-tests or Mann-Whitney U tests. Given our ablation framework, these pairwise evaluations are treated as \textit{a priori} planned contrasts~\cite{wickens2004design}; consequently, we report unadjusted $p$-values to mitigate the Type II error inflation inherent in conservative post-hoc corrections like Bonferroni.

\subsection{Implementation Details}

Narrative generation for all evaluated frameworks utilized \textit{gpt-5-mini-2025-08-07}. Textual evaluation via the LLM-as-a-Judge protocol used three open-weight models (Gemma 3 12B, Mistral 7B v0.3, and Qwen 3 14B). For visual assets, images were generated using the \textit{waiIllustriousSDXL} checkpoint in ComfyUI, while the VLM-guided generation loop and visual alignment evaluation utilized \textit{qwen3-vl:32b-instruct}. Generated scripts and multimedia assets were compiled using the Ren'Py engine. All additional materials, including the supplementary file, source code, data, prompts, and results, are available in the project GitHub repository.\footnote{\url{https://github.com/Yi-Xia-2010/SENSE-ToG}}

\section{Results}
\subsection{Framework Robustness and Complexity: Narrative Route Yield}
\label{subsec:route_yield}

Evaluating narrative generation frameworks requires ensuring the executability and structural integrity of multi-branching VNs. Beyond robustness, generating complex, intersecting narrative routes reflects each framework's structural richness and narrative complexity, quantified using the parsing protocol described in Section~\ref{subsub:Comparative_Integrity}. We define \textit{Narrative Route Yield} as the total number of distinct narrative routes produced by each framework. Table~\ref{tab:route_yield} summarizes the results, showing substantial differences in both robustness and narrative complexity among the tested frameworks: Full SENSE, its ablated variant withoutMIND, and an external Baseline. For complete transparency, the exhaustive route statistics, along with the raw generated Ren'Py script files, are provided in the project repository.

\begin{table}[tbp]
\centering
\caption{Comparison of Effective Narrative Route Yield by Genre.}
\label{tab:route_yield}
\begin{tabular}{lccc}
\hline
\textbf{Narrative Genre} & Baseline & WithoutMIND & Full SENSE \\ \hline
Historical & $12$ & $30$ & $\mathbf{75}$ \\
Romance & $48$ & $39$ & $\mathbf{71}$ \\
Sci-Fi & $39$ & $25$ & $\mathbf{80}$ \\ \hline
\textbf{Total Yield} & $99$ & $94$ & $\mathbf{226}$ \\ \hline
\end{tabular}
\end{table}

The probabilistic Baseline framework produced only 99 narrative routes, due to inherent limitations in its generation process combined with errors in the LLM-generated Ren'Py scripts. The ablated WithoutMIND framework yielded slightly fewer routes, generating only 94. This reduction stems from two main factors. First, the ablation method removes certain cycles and isolated nodes, which could have led to logically inconsistent routes during subsequent generation. Second, assigning a single emotional trajectory inherently limits route diversity. In contrast, Full SENSE framework generated 226 narrative routes, representing a more than 2.2-fold increase over Baseline. This improvement primarily reflects its ability to handle multiple emotional trajectories, together with the integrated \textit{Path-Aware Context} mechanism and \textit{Schr\"odinger Rule} for managing narrative convergences. 

\subsection{Narrative Quality}
\label{subsec:narrative_quality}

Using the Top-3 protocol described in Section~\ref{subsec: judge_protocol}, both WithoutMIND and Full SENSE frameworks generated all 54 evaluable routes, whereas the probabilistic Baseline produced only 43 due to generation and script errors, reflecting its structural and script limitations. Table~\ref{tab:main_results} reports the aggregated Top-3 scores, with comprehensive statistical details (including 95\% CIs, standard deviations, and individual judge $p$-values) provided in the supplementary file in the project repository. 

Crucially, Full SENSE achieved the highest Emotional Fidelity (EM: $M = 4.58$, 95\% CI $[4.53, 4.63]^{1,2}$) and Overall Quality (OV: $M = 4.52$, 95\% CI $[4.48, 4.56]^{2}$), significantly outperforming both Baseline and WithoutMIND configurations. RE and CH differences were not statistically significant, though WithoutMIND and Full SENSE consistently exceeded Baseline in CH. The Baseline reached the highest per-route Complexity (CX: $M = 4.74^{2,3\dagger}$, 95\% CI $[4.64, 4.84]$) but suffered from limited route coverage. WithoutMIND dropped to $M = 4.27$ (95\% CI $[4.17, 4.37]$) due to structural restrictions and single-emotion assignment, whereas Full SENSE successfully restored Complexity ($M = 4.53^{2}$, 95\% CI $[4.41, 4.65]$) and Surprise ($M = 3.56^{2}$, 95\% CI $[3.43, 3.69]$) through its multi-branch, multi-emotion handling mechanism. This observation is further supported by the AllMean robustness analysis (supplementary file in the project repository, Section S1.3), which evaluates the complete set of generated routes for each visual novel rather than only its Top-3 ranked routes. Despite the substantially larger evaluation set (226 routes), Full SENSE maintained high average narrative quality across all generated routes.

\begin{table}[tbp]
\centering
\caption{Top-3 Narrative Quality Comparison. Higher values indicate better performance.}
\label{tab:main_results}
\begin{threeparttable}
\resizebox{\columnwidth}{!}{
\begin{tabular}{l|cccccc|c}
\hline
\textbf{Method} & RE & CH & EM & SU & EG & CX & OV \\ \hline
(1) Baseline & 4.81 & 4.67 & 4.43 & 3.45 & 4.68 & \textcolor{ForestGreen}{\textbf{4.74$^{2,3\dagger}$}} & 4.47 \\
(2) WithoutMIND & \textcolor{ForestGreen}{\textbf{4.89}} & 4.72 & 4.48 & 3.31 & 4.75 & 4.27 & 4.42 \\
(3) Full SENSE & 4.85 & \textcolor{ForestGreen}{\textbf{4.75}} & \textcolor{ForestGreen}{\textbf{4.58$^{1,2}$}} & \textcolor{ForestGreen}{\textbf{3.56$^{2}$}} & \textcolor{ForestGreen}{\textbf{4.76}} & \textbf{4.53$^{2}$} & \textcolor{ForestGreen}{\textbf{4.52$^{2}$}} \\ \hline
\end{tabular}
}
\begin{tablenotes}
\item \textit{Note:} Values denote the macro-average scores across Gemma-3 (12B), Mistral (7B), and Qwen-3 (14B) evaluators. To maintain readability, standard deviations, 95\% Confidence Intervals (CIs), and effect sizes (Cohen's $d$) are exhaustively detailed in the supplementary file in the project repository.
\item \textit{Statistical Significance:} Superscripts indicate a statistically significant improvement ($p < 0.05$) over the corresponding method number. Significance was confirmed by a majority consensus (at least 2 of 3 judges) using independent Mann-Whitney U tests, strictly protected by a prior significant Kruskal-Wallis omnibus test to guard against FWER inflation.
\item $^\dagger$ The significantly higher CX score in Baseline is an artifact of unconstrained logical hallucinations rather than coherent narrative depth (see Sec.~\ref{sec:discussion}).
\end{tablenotes}
\end{threeparttable}
\end{table}

\subsection{Affective Salience and Narrative Pacing Results}
\label{subsec:affective_results}

Using the sentence-level valence ($V$) and arousal ($A$) predictions from the DeBERTa-based model described in Section~\ref{subsec:second_eval}, we computed route-level \textit{Affective Salience (AS)} and \textit{High-Salience Proportion (HSP)}. These metrics measure how frequently and strongly narrative sentences escape the neutral range $[0.35, 0.65]$, providing an objective assessment of emotional dynamics and narrative pacing (shown in Table~\ref{tab:affective_stats}).

\begin{table}[tbp]
\centering
\small 
\renewcommand{\arraystretch}{0.9} 
\setlength{\tabcolsep}{3pt} 
\caption{Route-level Affective Metrics (Mean and 95\% CI). Higher values indicate stronger affective salience.}
\label{tab:affective_stats}

\begin{tabular}{l cc cc}
\toprule
\multirow{2}{*}{\textbf{Method}} & \multicolumn{2}{c}{\textbf{Valence (Polarity)}} & \multicolumn{2}{c}{\textbf{Arousal (Intensity)}} \\ 
\cmidrule(lr){2-3} \cmidrule(lr){4-5}
 & \textbf{AS} & \textbf{HSP (\%)} & \textbf{AS} & \textbf{HSP (\%)} \\ 
\midrule
(1) Baseline & 0.084 & 15.3 & \textcolor{ForestGreen}{\textbf{0.246}$^{2}$} & \textcolor{ForestGreen}{\textbf{86.9}$^{2,3}$} \\
 & \scriptsize [0.082, 0.086] & \scriptsize [14.3, 16.3] & \scriptsize [0.244, 0.248] & \scriptsize [86.1, 87.6] \\
\addlinespace[3pt]

(2) WithoutMIND & 0.096$^{1}$ & 21.0$^{1}$ & 0.241 & 84.6 \\
 & \scriptsize [0.093, 0.099] & \scriptsize [19.8, 22.1] & \scriptsize [0.239, 0.244] & \scriptsize [83.9, 85.3] \\
\addlinespace[3pt]

(3) Full SENSE & \textcolor{ForestGreen}{\textbf{0.103}$^{1,2}$ }& \textcolor{ForestGreen}{\textbf{23.6}$^{1,2}$} & 0.244 & 85.0 \\
 & \scriptsize [0.101, 0.105] & \scriptsize [22.8, 24.4] & \scriptsize [0.242, 0.245] & \scriptsize [84.6, 85.5] \\ 
\bottomrule
\end{tabular}

\begin{tablenotes}
\footnotesize 
\item \textit{Note:} Brackets denote 95\% CI. Values in \textcolor{ForestGreen}{\textbf{bold green}} indicate the highest mean score in each column. Superscripts indicate statistically significant improvement ($p<0.05$) over the corresponding method number via Mann-Whitney U tests. Detailed test statistics and Cohen's $d$ effect sizes are provided in the supplementary file in the project repository.
\end{tablenotes}
\end{table}

\textbf{Valence Salience:} Full SENSE framework achieved the highest Valence AS ($M = 0.103$, 95\% CI $[0.101, 0.105]$) and HSP ($M = 23.6\%$, 95\% CI $[22.8, 24.4]$), significantly exceeding both Baseline and WithoutMIND ($p<0.001$). In contrast, Baseline narratives exhibited low emotional variation (Valence HSP $15.3\%$). The increase from WithoutMIND to Full SENSE indicates that enabling multiple intersecting trajectories allows higher affective salience without reducing structural diversity. Note that the 95\% CIs between Baseline and Full SENSE do not overlap, providing robust visual confirmation of statistical significance.

\textbf{Arousal Salience:} In the arousal dimension, the scores remained consistent with the base model's inherent adjustments and were not significantly affected by the proposed framework. No statistically significant difference in Arousal AS was observed between Full SENSE and Baseline ($p = 0.105$). While Baseline exhibited a slightly higher Arousal HSP ($86.9\%$) than Full SENSE ($85.0\%$, $p < 0.001$).

\subsection{Visual Evaluation Results}
\label{subsec:visual_evaluation_results}

Following Section~\ref{subsec:visual_evaluation}, we assessed semantic alignment (\textit{VQAScore}) across 23,291 total generated images (12,686 for Full SENSE; 7,405 for WithoutMIND; 3,200 for Baseline). As Table~\ref{tab:vqa_results} shows, Full SENSE achieved a character VQAScore of 0.862, showing no significant difference from WithoutMIND ($p = 0.545$) but significantly outperforming Baseline ($p < 0.001$), whereas WithoutMIND scored higher on backgrounds ($p < 0.001$). One possible speculation is that MIND incorporates additional affective guidance, shifting the focus of image-generation prompts from concrete object descriptions to abstract atmospheric elements. Additionally, we evaluated character identity preservation (CLIP-I, DINO) specifically on the character asset subsets (7,981 for Full SENSE; 3,855 for WithoutMIND; 1,306 for Baseline). Full SENSE achieved the highest CLIP-I and DINO. It significantly outperformed Baseline ($p < 0.001$). Compared to WithoutMIND, Full SENSE demonstrated significant improvement in identity retention as measured by DINO ($p < 0.001$) while performing comparably in CLIP-I ($p = 0.414$). This confirms SENSE effectively retains character consistency even across diverse emotional states.

\begin{table}[tbp]
\centering
\caption{Visual Evaluation Results Across Frameworks (Mean). Higher values indicate better performance.}
\label{tab:vqa_results}
\footnotesize 
\renewcommand{\arraystretch}{0.95}
\setlength{\tabcolsep}{4pt} 
\begin{tabular}{lccc}
\toprule
\textbf{Metric / Category} & (1) Baseline & (2) WithoutMIND & (3) Full SENSE \\ 
\midrule
\multicolumn{4}{l}{\textit{Semantic Alignment (VQAScore $\uparrow$):}} \\
Backgrounds & 0.590 & \textcolor{ForestGreen}{\textbf{0.750}$^{1,3}$} & 0.672$^{1}$ \\
Characters & 0.601 & \textcolor{ForestGreen}{\textbf{0.862}$^{1}$} & \textcolor{ForestGreen}{\textbf{0.862}$^{1}$} \\ 
Total Mean Score & 0.595 & \textcolor{ForestGreen}{\textbf{0.815}$^{1,3}$} & 0.803$^{1}$ \\
\midrule
\multicolumn{4}{l}{\textit{Character Identity Preservation $\uparrow$:}} \\
CLIP-I & 0.824 & 0.942$^{1}$ & \textcolor{ForestGreen}{\textbf{0.958}$^{1}$} \\
DINO & 0.611 & 0.933$^{1}$ & \textcolor{ForestGreen}{\textbf{0.972}$^{1,2}$} \\
\bottomrule
\end{tabular}
% ------------------------------------------------
\begin{tablenotes}
\footnotesize
\item \textit{Note:} Values in \textcolor{ForestGreen}{\textbf{bold green}} indicate the highest mean score in each row. Superscripts indicate statistically significant improvement ($p<0.05$) over the corresponding method number via planned ablation contrasts (Mann-Whitney U tests). Detailed test statistics and 95\% confidence intervals (CIs) are provided in the supplementary file in the project repository.
\end{tablenotes}
\end{table}
\vspace{-3mm} 

\subsection{End-to-End System Deployment}
\label{subsec:system_deployment}

\begin{figure}[tbp]
    \centering
    \includegraphics[width=\linewidth]{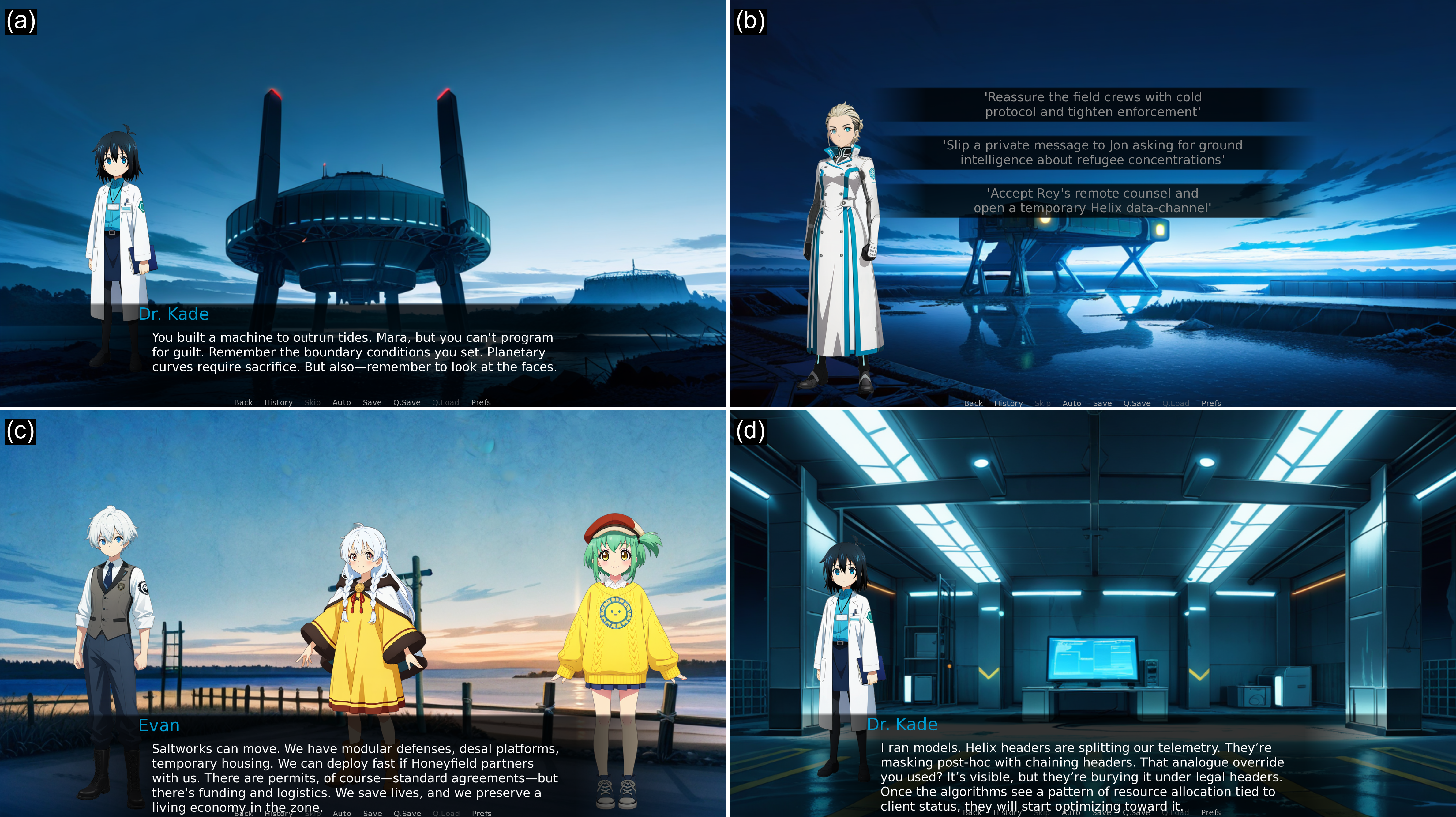}
    \caption{Gameplay captures of the SENSE-generated VNs. The subfigures illustrate 
    (a) integration of assets within a sci-fi setting; 
    (b) interactive choice menus; 
    (c) multi-asset integration in a romance narrative; 
    (d) the same character sprite in a different background.}
    \label{fig:gameplay_full_results}
\end{figure}

We deployed the SENSE framework as a fully playable VNs using the Ren'Py engine. Figure~\ref{fig:gameplay_full_results} shows representative gameplay captures. The system executes the full pipeline without manual intervention. It demonstrates robust integration of narrative, visual, and interactive components, while preserving semantic alignment between characters, backgrounds, and story events.

\subsection{User Study}
To complement the automatic evaluation, we conducted a user study using a representative Romance scenario, specifically a median-performing case (ranked third among the six generated Romance narratives based on LLM-as-Judge scores), given the genre's widespread adoption in visual novels. As the baseline-generated Ren'Py project was not directly executable, we reimplemented the Ren’Py project and conducted script modifications required for interactive evaluation while preserving the original generated narrative and branching structure. Prior to the study, informed consent was obtained from all participants. A total of 20 participants completed the study. Two participants were excluded because their reading speeds were identified as extreme outliers, suggesting that the visual novel content was unlikely to have been read attentively, leaving a final sample of 18 participants for analysis (mean age = 23.2 years, SD = 2.9). To establish baseline familiarity with the medium, participants also self-reported their prior visual novel experience on a 7-point scale.

Participants experienced both versions in randomized order. After each session, they completed a post-session questionnaire assessing enjoyment, boredom, theme relevance, genre relevance, image consistency, and narrative coherence on a 7-point Likert scale, with the boredom item reverse-scored during statistical analysis. After experiencing both versions, participants additionally identified their preferred version, the version that they perceived as having a more perceptible emotional arc, and provided qualitative feedback. Additional implementation and study details are provided in the supplementary file in the project repository.

To robustly handle discrete ties without relying on rank-based tests, we analyzed responses using pairwise bootstrap resampling ($n = 10,000$) across two groups: the overall sample ($N=18$) and a subgroup of experienced players (experience score $\ge 5$, $N=8$). Detailed mean differences and 95\% confidence intervals are provided in Table \ref{tab:bootstrap_results}. Unadjusted $p$-values were used not only because these are planned comparisons, but also because strict corrections would severely inflate Type II errors in a small sample size. Overall, participants rated our method higher in Genre Relevance ($M_{diff} = 0.67, p = 0.013$), with marginal trends in Image Consistency ($M_{diff} = 0.72, p = 0.053$) and Narrative Coherence ($M_{diff} = 0.56, p = 0.078$). Crucially, the experienced subgroup reported significantly higher Theme Relevance ($M_{diff} = 1.13, p < 0.001$), suggesting that greater player experience heightens sensitivity to thematic alignment. Furthermore, in a forced-choice assessment, 66.7\% ($12/18$) of participants identified our method as having a clearer emotional arc, while preferences for overall enjoyment were evenly split ($9$ vs. $9$).

\begin{table}[tbp]
\centering
\caption{Pairwise Bootstrap Resampling Results ($M_{diff}$ and 95\% CI)}
\label{tab:bootstrap_results}
\resizebox{\columnwidth}{!}{%
\begin{tabular}{@{}lcc@{}}
\toprule
\textbf{Metric} & \textbf{Overall ($N=18$)} & \textbf{Exp. ($n=8$)} \\
\midrule
Enjoyment & $0.56\ [-0.22, 1.39]$ & $0.38\ [-0.50, 1.50]$ \\
Boredom (Rev.) & $-0.11\ [-0.67, 0.44]$ & $-0.25\ [-1.13, 0.63]$ \\
Theme Relevance & $0.50\ [-0.44, 1.39]$ & $\mathbf{1.13\ [0.25, 2.25]}^{***}$ \\
Genre Relevance & $\mathbf{0.67\ [0.11, 1.39]}^{*}$ & $0.38\ [-0.13, 0.88]^{\dagger}$ \\
Image Consistency & $0.72\ [-0.06, 1.56]^{\dagger}$ & $0.63\ [-0.25, 1.63]$ \\
Narrative Coherence & $0.56\ [-0.11, 1.28]^{\dagger}$ & $0.25\ [-0.38, 1.00]$ \\
\bottomrule
\multicolumn{3}{@{}l@{}}{\small \textit{Note:} $M_{diff}$ is the mean difference. `Exp.' denotes players with VN} \\
\multicolumn{3}{@{}l@{}}{\small experience $\ge 5$. $^{\dagger} p < 0.10$, $^{*} p < 0.05$, $^{***} p < 0.001$ (unadjusted).} \\
\end{tabular}%
}
\end{table}

\subsection{Generalizability to Open-Weight Models}
\label{subsec:open_weight_evaluation}

To evaluate framework generalizability, we replicated our primary experiments using the open-weight qwen3.6:35b-a3b-bf16 model~\cite{qwen36_35b_a3b}, deferring comprehensive statistical details to the supplementary file in the project repository. Full SENSE generated 111 distinct narrative routes, substantially outperforming both WithoutMIND ($N=53$) and Baseline ($N=37$), suggesting enhanced generation for models with strong instruction-following capabilities. While Full SENSE and WithoutMIND achieved higher absolute OV scores (Table~\ref{tab:qwen_summary}), these improvements did not reach strict statistical consensus.

In terms of affective and visual performance, both Full SENSE configurations achieved significantly higher Valence HSP compared to Baseline ($p < 0.001$). For visual evaluation, while WithoutMIND scored highest in global text-image alignment, Full SENSE achieved the highest character identity preservation (CLIP-I, DINO), aligning with our primary results and demonstrating the image generation pipeline's resilience when utilizing sufficiently capable models.

\begin{table}[tpb]
\centering
\caption{Summary of Qwen-3.6 Case Study Performance. Higher values indicate better performance.}
\label{tab:qwen_summary}
\resizebox{\columnwidth}{!}{%
\renewcommand{\arraystretch}{1.1}
\setlength{\tabcolsep}{4pt}
\begin{tabular}{lccc}
\toprule
\textbf{Metric} & (1) Baseline & (2) WithoutMIND & (3) Full SENSE \\ 
\midrule
Total Route Yield & 37 & 53 & \textcolor{ForestGreen}{\textbf{111}} \\
Overall Quality (OV) & 4.50 & 4.58 & \textcolor{ForestGreen}{\textbf{4.61}} \\
Valence AS & 0.130 & 0.158$^{1}$ & \textcolor{ForestGreen}{\textbf{0.162}$^{1}$} \\
Valence HSP (\%) & 34.2 & 46.5$^{1}$ & \textcolor{ForestGreen}{\textbf{47.0}$^{1}$ }\\
VQAScore (Global) & 0.778 & \textcolor{ForestGreen}{\textbf{0.825}$^{1,3}$} & 0.805$^{1}$ \\
CLIP-I & 0.846 & 0.937$^{1}$ & \textcolor{ForestGreen}{\textbf{0.951}$^{1,2}$} \\
DINO & 0.438 & 0.924$^{1}$ & \textcolor{ForestGreen}{\textbf{0.942}$^{1,2}$} \\
\bottomrule
\end{tabular}%
}
\begin{tablenotes}
\footnotesize
\item \textit{Note:} Values denote absolute means, with the best performance in each row highlighted in \textcolor{ForestGreen}{\textbf{bold}}. Superscripts indicate a statistically significant improvement ($p < 0.05$) over the corresponding method number. For Overall Quality (OV), differences did not reach strict statistical consensus across evaluators; thus, no significance superscripts are applied. Detailed statistics are available in the supplementary file in the project repository.
\end{tablenotes}
\end{table}

\section{Discussion}
\label{sec:discussion}

\subsection{Valence Guidance and Multi-Track Emotional Support}

The Top-3 LLM evaluation results (Section~\ref{subsec:narrative_quality}) show that Full SENSE framework achieves the highest Empathy (EM: $4.58^{1,2}$) and Overall Quality (OV: $4.52^{2}$) among all tested methods, with key comparisons reaching statistical significance (as indicated by superscripts). Compared with Baseline and WithoutMIND, Full SENSE achieves higher emotional engagement and expressive depth according to the LLM judges. The use of a single emotional arc in WithoutMIND constrains the diversity of emotional trajectories across routes and may, to a certain extent, limit the perceived emotional richness of the generated VNs. In contrast, MIND enables multiple simultaneous emotional tracks, supporting the generation of narratives in which emotional arcs are both structurally coherent and affectively varied.

Sentence-level analysis of Valence Affective Salience (AS) and High-Salience Proportion (HSP) (Section~\ref{subsec:affective_results}) provides complementary quantitative support. Full SENSE achieves the highest Valence AS ($0.103$) and HSP ($23.6\%$), significantly exceeding both Baseline ($15.3\%$) and WithoutMIND ($21.0\%$). In contrast, differences in Arousal AS and HSP are minor across frameworks (Full SENSE: $0.244$, $85.0\%$; Baseline: $0.246$, $86.9\%$). This pattern aligns with classic computational narrative analysis \cite{reagan2016emotional}, where ``emotional arcs'' are primarily quantified by shifts in narrative polarity (valence). Since LLMs internalize these structural conventions during pretraining, they inherently interpret pure ``Rise'' and ``Fall'' directives as valence shifts. As a result, while SENSE successfully modulates valence across branching narratives, overall arousal levels remain largely driven by the baseline language model's inherent generative tendencies. 

\subsection{The Complexity Paradox: Global vs. Per-Route Narrative Complexity}

Our results reveal a complexity paradox inherent to branching VNs. Full SENSE generates a large number of routes (226), which exponentially increases the challenge of maintaining structural integrity and emotional coherence per path. In contrast, Baseline framework produces significantly fewer routes (99); this allows it to attain a relatively high per-route complexity (CX), but at the severe expense of overall narrative coverage, branching richness, and emotional diversity. This trade-off demonstrates that global narrative complexity imposes strict constraints on local coherence. However, by leveraging SENSE’s MIND cognitive module,  structure analyzer, and path-aware context management, our framework successfully overcomes this paradox. It effectively balances macro-level structural richness with micro-level coherence, maintaining a robust level of logical and affective integrity across a vastly expanded narrative space.

\subsection{User Study Findings}
\label{subsec:user_study_findings}
Although overall enjoyment was evenly split (9 vs. 9), the broader pattern of results points toward an advantage for our framework. Participants rated our framework higher in Genre Relevance, with marginal trends in Image Consistency and Narrative Coherence pointing in the same direction. Theme Relevance showed the largest mean difference among all metrics, reaching significance within the experienced subgroup ($n=8$, $p<0.001$) though not in the overall sample, indicating that participants with greater prior VN experience tended to rate our framework as more thematically relevant than the baseline. Consistent with these results, 66.7\% of participants identified our version as having a clearer emotional arc, echoing comments describing a ``more emotional" story with a ``better representation of the given theme." The baseline, meanwhile, was criticized for visual anomalies described as ``inconsistent AI slop," including sprites ``stacked on top of another image," aligning with its lower ratings on Image Consistency.

This raises the question of why enjoyment itself did not follow the same pattern, a tension partly clarified by the qualitative feedback: some players found the baseline's ``chaotic" and ``haphazard" transitions entertaining because of their unpredictability, whereas the visual stability of our framework was, for some, described as ``tedious" and lacking variation by comparison. This suggests that enjoyment in visual novels may be shaped not only by narrative or thematic strength, but also by visual variety and unpredictability, an interaction that merits further investigation in future work. Dedicated players also noted prose and character-writing limitations shared across both versions, including ``unnatural" text lacking ``proper character introduction," and characters described as feeling ``empty and bland" without clear motivation or background. These observations point to text quality and character depth as an area for improvement in future iterations. 

\subsection{Limitations and Future Work}
\label{sec:limitations}

Despite SENSE's robust performance, several limitations remain. First, while the framework successfully manages complex branching structures across different models (yielding highly competitive results even on open-weight models like Qwen-3.6), there remains a noticeable gap in literary quality between LLM-generated text and authentic, human-authored visual novels. Current models often struggle to produce the nuanced prose and deep character motivations expected in commercial VNs, meaning the system's ultimate narrative depth is inherently bounded by the LLM's creative limitations. Second, although the revised pipeline successfully incorporates facial expressions while preserving character identity, modeling expressive body poses and gestures remains an orthogonal challenge, as current image models still struggle to preserve character identity consistency across diverse poses. Third, while our framework effectively modulates narrative valence, achieving fine-grained guidance over narrative arousal remains limited by the LLM's inherent tendencies. Finally, our user study ($N=18$) provides suggestive trends that warrant validation in broader contexts.

Future work will explore pose-constrained image generation and context-aware inpainting to improve cross-prompt visual consistency for dynamic body gestures. Additionally, developing more effective arousal guidance mechanisms, enhancing the literary depth of generated prose (e.g., via specialized fine-tuning), and conducting larger participatory user studies with multimodal physiological metrics \cite{gursesli2026multimodal} are critical next steps to better evaluate player emotional engagement.

\subsection{Ethical Considerations}
\label{sec:ethics}
The use of generative AI in creative industries raises ethical concerns, particularly regarding potential displacement of human artists. SENSE is designed not to replace creators, but to serve as a ``Human-AI Co-creation" tool. By automating laborious branching logic and repetitive asset generation, it lowers the barrier for indie developers and small studios. The SENSE generates an initial, emotionally rich, playable VN that serves as a baseline reference for creators. This AI-constructed “affective draft” acts as a catalyst, allowing human creators to further refine the emotional nuances and expand the narrative.

Additionally, because SENSE’s emotion navigation module can generate routes involving negative emotions or extreme conflict, there is an inherent risk of producing harmful, biased, or inappropriate content. For industrial deployment, it is essential to implement stricter safety alignment and robust content filtering on top of SENSE, ensuring that automatically generated interactive narratives and multimedia assets adhere to ethical standards and age-appropriate guidelines.

\section{Conclusion}
We presented SENSE, a state-aware framework for generating playable, branching VNs with multi-track emotional navigation from minimal user input. By integrating the MIND cognitive module, a structure analyzer, and path-aware context management, SENSE produces narratives that are both structurally coherent and emotionally resonant. Extensive evaluations demonstrate significant improvements in emotional richness, narrative diversity, and the visual-textual alignment of generated assets, all while maintaining a robust end-to-end integration within a deterministic game environment.

Overall, SENSE demonstrates the potential of AI-assisted tools to accelerate Human-AI co-creation, allowing creators to focus on high-level narrative design, artistic refinement, and emotionally engaging storytelling in playable interactive experiences.

\bibliography{ref}

@article{reagan2016emotional,
  title={The emotional arcs of stories are dominated by six basic shapes},
  author={Reagan, Andrew J and Mitchell, Lewis and Kiley, Dilan and Danforth, Christopher M and Dodds, Peter Sheridan},
  journal={EPJ data science},
  volume={5},
  number={1},
  pages={31},
  year={2016},
  publisher={Springer}
}

@article{wen2025all,
  title={All stories are one story: Emotional arc guided procedural game level generation},
  author={Wen, Yunge and Huang, Chenliang and Zhou, Hangyu and Zeng, Zhuo and Po, Chun Ming Louis and Togelius, Julian and Merino, Timothy and Earle, Sam},
  journal={arXiv preprint arXiv:2508.02132},
  year={2025}
}

@article{gursesli2026multimodal,
  title={Multimodal Analysis of Emotions in Gaming: Understanding Cultural Influences},
  author={Gursesli, Mustafa Can and Tarchi, Pietro and Cal{\'a}, Federico and Frassineti, Lorenzo and Guazzini, Andrea and Duradoni, Mirko and Park, Kyoungju and Thawonmas, Ruck and You, Xiao and Lanata, Antonio},
  journal={IEEE Transactions on Affective Computing},
  year={2026},
  publisher={IEEE}
}

@inproceedings{10.1145/3746027.3755541,
author = {Zhang, Yilin and Wei, Yanyan and Zhang, Zhao and Fan, Jicong and Zhang, Haijun and Yan, Shuicheng},
title = {From Outline to Detail: An Hierarchical End-to-end Framework for Coherent and Consistent Visual Novel Generation and Assembly},
year = {2025},
isbn = {9798400720352},
publisher = {Association for Computing Machinery},
address = {New York, NY, USA},
url = {https://doi.org/10.1145/3746027.3755541},
doi = {10.1145/3746027.3755541},
booktitle = {Proceedings of the 33rd ACM International Conference on Multimedia},
pages = {8506–8516},
numpages = {11},
location = {Dublin, Ireland},
series = {MM '25}
}

@Article{sym18010113,
AUTHOR = {Puchalski, Marcin and Woźna-Szcześniak, Bożena},
TITLE = {Symmetry-Aware LLM-Driven Generation and Repair of Interactive Fiction Graphs in Twine/Twee},
JOURNAL = {Symmetry},
VOLUME = {18},
YEAR = {2026},
NUMBER = {1},
ARTICLE-NUMBER = {113},
URL = {https://www.mdpi.com/2073-8994/18/1/113},
ISSN = {2073-8994},
DOI = {10.3390/sym18010113}
}

@article{revers2025dissociating,
  title={Dissociating affective salience and valence in (very) long-latency ERPs},
  author={Revers, Hans and Stekelenburg, Jeroen J and Vroomen, Jean and Van Deun, Katrijn and Bastiaansen, Marcel},
  journal={Psychophysiology},
  volume={62},
  number={3},
  pages={e70030},
  year={2025},
  publisher={Wiley Online Library}
}

@article{Gursesli16022026,
    author = {Mustafa Can Gursesli and Siyuan Chen and Mury Fajar Dewantoro and Xiao You and Ege Anbar and Pittawat Taveekitworachai and Febri Abdullah and Pietro Tarchi and Mirko Duradoni and Antonio Lanata and Andrea Guazzini and Ruck Thawonmas},
    title = {The Role of Large Language Model-Generated Stories in the Narrative Experience of Serious Visual Novel Games},
    journal = {International Journal of Human–Computer Interaction},
    volume = {42},
    number = {4},
    pages = {2489--2508},
    year = {2026},
    publisher = {Taylor \& Francis},
    doi = {10.1080/10447318.2025.2528993},
    URL = {https://doi.org/10.1080/10447318.2025.2528993},
    eprint = {https://doi.org/10.1080/10447318.2025.2528993}
}

@article{10.1162/tacl_a_00689,
    author = {Chhun, Cyril and Suchanek, Fabian M. and Clavel, Chloé},
    title = {Do Language Models Enjoy Their Own Stories? Prompting Large Language Models for Automatic Story Evaluation},
    journal = {Transactions of the Association for Computational Linguistics},
    volume = {12},
    pages = {1122-1142},
    year = {2024},
    month = {09},
    issn = {2307-387X},
    doi = {10.1162/tacl_a_00689},
    url = {https://doi.org/10.1162/tacl_a_00689},
    eprint = {https://direct.mit.edu/tacl/article-pdf/doi/10.1162/tacl_a_00689/2470807/tacl_a_00689.pdf},
}

@inproceedings{christ-etal-2024-modeling,
    title = "Modeling Emotional Trajectories in Written Stories Utilizing Transformers and Weakly-Supervised Learning",
    author = {Christ, Lukas  and
      Amiriparian, Shahin  and
      Milling, Manuel  and
      Aslan, Ilhan  and
      Schuller, Bj{\"o}rn},
    editor = "Ku, Lun-Wei  and
      Martins, Andre  and
      Srikumar, Vivek",
    booktitle = "Findings of the Association for Computational Linguistics: ACL 2024",
    month = aug,
    year = "2024",
    address = "Bangkok, Thailand",
    publisher = "Association for Computational Linguistics",
    url = "https://aclanthology.org/2024.findings-acl.426/",
    doi = "10.18653/v1/2024.findings-acl.426",
    pages = "7144--7159"
}

@InProceedings{10.1007/978-3-031-72673-6_20,
author="Lin, Zhiqiu
and Pathak, Deepak
and Li, Baiqi
and Li, Jiayao
and Xia, Xide
and Neubig, Graham
and Zhang, Pengchuan
and Ramanan, Deva",
editor="Leonardis, Ale{\v{s}}
and Ricci, Elisa
and Roth, Stefan
and Russakovsky, Olga
and Sattler, Torsten
and Varol, G{\"u}l",
title="Evaluating Text-to-Visual Generation with Image-to-Text Generation",
booktitle="Computer Vision -- ECCV 2024",
year="2025",
publisher="Springer Nature Switzerland",
address="Cham",
pages="366--384",
isbn="978-3-031-72673-6"
}

@book{wickens2004design,
  title={Design and analysis: A researcher's handbook},
  author={Wickens, Thomas D and Keppel, Geoffrey},
  volume={860},
  year={2004},
  publisher={Pearson Prentice-Hall Upper Saddle River, NJ}
}

@inproceedings{park2023generative,
  author    = {Park, Joon Sung and O'Brien, Joseph C. and Cai, Carrie J. and Morris, Meredith Ringel and Liang, Percy and Bernstein, Michael S.},
  title     = {Generative Agents: Interactive Simulacra of Human Behavior},
  booktitle = {Proceedings of the 36th Annual ACM Symposium on User Interface Software and Technology},
  year      = {2023}
}

@inproceedings{
wu2024autogen,
title={AutoGen: Enabling Next-Gen {LLM} Applications via Multi-Agent Conversation},
author={Qingyun Wu and Gagan Bansal and Jieyu Zhang and Yiran Wu and Beibin Li and Erkang Zhu and Li Jiang and Xiaoyun Zhang and Shaokun Zhang and Jiale Liu and Ahmed Hassan Awadallah and Ryen W White and Doug Burger and Chi Wang},
booktitle={ICLR 2024 Workshop on Large Language Model (LLM) Agents},
year={2024},
url={https://openreview.net/forum?id=uAjxFFing2}
}

@inproceedings{hong2024metagpt,
  author    = {Hong, Sirui and Zhuge, Mingchen and Chen, Jonathan and Zheng, Xiawu and Cheng, Yuheng and Wang, Jinlin and Zhang, Ceyao and Wang, Zili and Yau, Steven Ka Shing and Lin, Zijuan and Zhou, Liyang and Ran, Chenyu and Xiao, Lingfeng and Wu, Chenglin and Schmidhuber, J{\"u}rgen},
  title     = {MetaGPT: Meta Programming for A Multi-Agent Collaborative Framework},
  booktitle = {International Conference on Learning Representations},
  year      = {2024}
}

@inproceedings{liang2024encouraging,
  author    = {Liang, Tian and He, Zhiwei and Jiao, Wenxiang and Wang, Xing and Wang, Yan and Wang, Rui and Yang, Yujiu and Tu, Zhaopeng and Shi, Shuming},
  title     = {Encouraging Divergent Thinking in Large Language Models through Multi-Agent Debate},
  booktitle = {Proceedings of the 2024 Conference on Empirical Methods in Natural Language Processing},
  year      = {2024}
}

@inproceedings{du2024improving,
  author    = {Du, Yilun and Li, Shuang and Torralba, Antonio and Tenenbaum, Joshua B. and Mordatch, Igor},
  title     = {Improving Factuality and Reasoning in Language Models through Multiagent Debate},
  booktitle = {Proceedings of the 41st International Conference on Machine Learning},
  year      = {2024}
}

@inproceedings{li2023camel,
  author    = {Li, Guohao and Hammoud, Hasan Abed Al Kader and Itani, Hani and Khizbullin, Dmitrii and Ghanem, Bernard},
  title     = {CAMEL: Communicative Agents for ``Mind'' Exploration of Large Language Model Society},
  booktitle = {Advances in Neural Information Processing Systems},
  volume    = {36},
  year      = {2023}
}

@inproceedings{wang2024ibsen,
  author    = {Han, Suhan and Chen, Lili and Lin, Lulu and Xu, Zihao and Yu, Kai},
  title     = {IBSEN: Director-Actor Agent Collaboration for Controllable and Interactive Drama Script Generation},
  booktitle = {Proceedings of the 62nd Annual Meeting of the Association for Computational Linguistics},
  pages     = {1607--1619},
  year      = {2024}
}

@inproceedings{micaloni2024agents,
  author    = {Huot, Fantine and Amplayo, Reinald Kim and Palomaki, Jennimaria and Jakobovits, Anna Googol and Clark, Elizabeth and Lapata, Mirella},
  title     = {Agents' Room: Narrative Generation through Multi-step Collaboration},
  booktitle = {The Thirteenth International Conference on Learning Representations},
  year      = {2025}
}

@inproceedings{zhang2025apbased,
  author    = {Zhang, Menghan and Ye, Zihao and Suntichaikul, Pujana and Goto, Shigeyuki and Nakamura, Osamu and Thawonmas, Ruck},
  title     = {AP-based LLM Story Generation: Envision Technology Development through Science Fiction},
  booktitle = {2025 IEEE 14th Global Conference on Consumer Electronics (GCCE)},
  pages     = {631--634},
  year      = {2025}
}

@inproceedings{delima2024patternteller,
  title={A pattern-oriented AI-powered approach to story composition},
  author={Lima, Edirlei Soares de and Neggers, Margot ME and Casanova, Marco A and Feij{\'o}, Bruno and Furtado, Antonio L},
  booktitle={International Conference on Entertainment Computing},
  pages={135--150},
  year={2024},
  organization={Springer}
}

@inproceedings{theune2003virtual,
  title={The virtual storyteller: Story creation by intelligent agents},
  author={Theune, Mari{\"e}t and Faas, Sander and Heylen, Dirk KJ and Nijholt, Antinus},
  booktitle={1st International Conference on Technologies for Interactive Digital Storytelling and Entertainment 2003},
  pages={204--215},
  year={2003},
  organization={Fraunhofer IRB Verlag}
}

@inproceedings{hernandez2015keeping,
  title={Keeping the player on an emotional trajectory in interactive storytelling},
  author={Hernandez, Sergio Poo and Bulitko, Vadim and Spetch, Marcia},
  booktitle={Proceedings of the AAAI conference on artificial intelligence and interactive digital entertainment},
  volume={11},
  number={1},
  pages={65--71},
  year={2015}
}

@inproceedings{meehan1977tale,
  title={TALE-SPIN, An Interactive Program that Writes Stories.},
  author={Meehan, James R},
  booktitle={Ijcai},
  volume={77},
  pages={91--98},
  year={1977}
}

@inproceedings{martin2018event,
  title={Event representations for automated story generation with deep neural nets},
  author={Martin, Lara and Ammanabrolu, Prithviraj and Wang, Xinyu and Hancock, William and Singh, Shruti and Harrison, Brent and Riedl, Mark},
  booktitle={Proceedings of the AAAI Conference on Artificial Intelligence},
  volume={32},
  number={1},
  year={2018}
}

@inproceedings{fan2018hierarchical,
  title={Hierarchical neural story generation},
  author={Fan, Angela and Lewis, Mike and Dauphin, Yann},
  booktitle={Proceedings of the 56th Annual Meeting of the Association for Computational Linguistics (Volume 1: Long Papers)},
  pages={889--898},
  year={2018}
}

@inproceedings{wang2025generating,
  title={Generating long-form story using dynamic hierarchical outlining with memory-enhancement},
  author={Wang, Qianyue and Hu, Jinwu and Li, Zhengping and Wang, Yufeng and Li, Daiyuan and Hu, Yu and Tan, Mingkui},
  booktitle={Proceedings of the 2025 Conference of the Nations of the Americas Chapter of the Association for Computational Linguistics: Human Language Technologies (Volume 1: Long Papers)},
  pages={1352--1391},
  year={2025}
}

@inproceedings{rashkin2020plotmachines,
  title={PlotMachines: Outline-conditioned generation with dynamic plot state tracking},
  author={Rashkin, Hannah and Celikyilmaz, Asli and Choi, Yejin and Gao, Jianfeng},
  booktitle={Proceedings of the 2020 Conference on Empirical Methods in Natural Language Processing (EMNLP)},
  pages={4274--4295},
  year={2020}
}

@article{huang2024if,
  title={WHAT-IF: Exploring branching narratives by meta-prompting large language models},
  author={Huang, Runsheng and Martin, Lara J and Callison-Burch, Chris and others},
  journal={arXiv preprint arXiv:2412.10582},
  year={2024}
}

@inproceedings{leandro2024geneva,
  title={GENEVA: GENErating and Visualizing branching narratives using LLMs},
  author={Leandro, Jorge and Rao, Sudha and Xu, Michael and Xu, Weijia and Jojic, Nebojsa and Brockett, Chris and Dolan, Bill},
  booktitle={2024 IEEE Conference on Games (CoG)},
  pages={1--5},
  year={2024},
  organization={IEEE}
}

@article{wang2025taleframe,
  title={TaleFrame: An Interactive Story Generation System with Fine-Grained Control and Large Language Models},
  author={Wang, Yunchao and Sun, Guodao and Fu, Zihang and Liu, Zhehao and Du, Kaixing and Gao, Haidong and Liang, Ronghua},
  journal={arXiv preprint arXiv:2512.02402},
  year={2025}
}

@inproceedings{rameshkumar2025reasoning,
  title={Reasoning Models Reason Well, Until They Don’t},
  author={Rameshkumar, Revanth and Huang, Jimson and Sun, Yunxin and Xia, Fei and Saparov, Abulhair},
  booktitle={Proceedings of the 14th International Joint Conference on Natural Language Processing and the 4th Conference of the Asia-Pacific Chapter of the Association for Computational Linguistics},
  pages={936--956},
  year={2025}
}

@article{ekman1992argument,
  title={An argument for basic emotions},
  author={Ekman, Paul},
  journal={Cognition \& emotion},
  volume={6},
  number={3-4},
  pages={169--200},
  year={1992},
  publisher={Taylor \& Francis}
}

@article{cowen2017self,
  title={Self-report captures 27 distinct categories of emotion bridged by continuous gradients},
  author={Cowen, Alan S and Keltner, Dacher},
  journal={Proceedings of the national academy of sciences},
  volume={114},
  number={38},
  pages={E7900--E7909},
  year={2017},
  publisher={National Academy of Sciences}
}

@article{ye2023ip-adapter,
  title={IP-Adapter: Text Compatible Image Prompt Adapter for Text-to-Image Diffusion Models},
  author={Ye, Hu and Zhang, Jun and Liu, Sibo and Han, Xiao and Yang, Wei},
  journal ={arXiv preprint arxiv:2308.06721},
  year={2023}
}

@software{yolov8_ultralytics,
      author = {Glenn Jocher and Ayush Chaurasia and Jing Qiu},
      title = {Ultralytics YOLOv8},
      version = {8.0.0},
      year = {2023},
      url = {https://github.com/ultralytics/ultralytics},
      orcid = {0000-0001-5950-6979, 0000-0002-7603-6750, 0000-0003-3783-7069},
      license = {AGPL-3.0}
    }

@inproceedings{ruiz2023dreambooth,
  title={Dreambooth: Fine tuning text-to-image diffusion models for subject-driven generation},
  author={Ruiz, Nataniel and Li, Yuanzhen and Jampani, Varun and Pritch, Yael and Rubinstein, Michael and Aberman, Kfir},
  booktitle={2023 IEEE/CVF Conference on Computer Vision and Pattern Recognition (CVPR)},
  pages={22500--22510},
  year={2023},
  organization={IEEE}
}

@misc{qwen36_35b_a3b,
    title = {{Qwen3.6-35B-A3B}: Agentic Coding Power, Now Open to All},
    url = {https://qwen.ai/blog?id=qwen3.6-35b-a3b},
    author = {{Qwen Team}},
    month = {April},
    year = {2026}
}
\bibliographystyle{IEEEtran}
\end{document}